\documentclass[aip,rsi,amsmath,amssymb]{revtex4-1} 
\usepackage{CJK}
\usepackage[english]{babel}
\usepackage[utf8x]{inputenc}
\usepackage{graphics}
\usepackage{amsmath}
\usepackage{graphicx}
\usepackage{setspace}
\usepackage{dcolumn}
\usepackage{bm}

\begin{document}

\title{The Saturated and Supercritical Stirling Cycle Thermodynamic Heat Engine Cycle}

\author{Matthew David Marko\\mattdmarko@gmail.com\\Orcid ID: 0000-0002-6775-7636\\Marko Motors LLC, Seaside Heights NJ 08751 USA}

\maketitle

\section*{Abstract}

\noindent On the assumption that experimentally validated tabulated thermodynamic properties of saturated fluids published by the National Institute of Standards and Technology are accurate, a theoretical thermodynamic cycle can be demonstrated that produces a net-negative entropy generation to the universe.  The experimental data on the internal energy can also be used to obtain a simple, empirical equation for the change in internal energy of a real fluid undergoing isothermal expansion and compression.  This demonstration provides experimental evidence to the theory that temperature-dependent intermolecular attractive forces can be an entropic force that can enhance the thermodynamic efficiency of a real-fluid macroscopic heat engine to exceed that of the Carnot efficiency.

\section{Introduction}

In the design of any thermodynamic system to convert heat to and from mechanical work, the laws of thermodynamics must always be considered.  The first law of thermodynamics states that energy can not be created or destroyed, and that the change in internal energy equals the heat and work input into the working fluid \cite{ClausiusOrig,1,2,3,4,StatThermo}
\begin{eqnarray}
\label{eq:eqFirstLaw}
{{\delta}u}&=&{{\delta}q}-{{\delta}w},
\end{eqnarray}
where ${{\delta}u}$ (J/kg) is the change in specific internal energy, ${\delta}q$ (J/kg) is the specific heat transfered, and \emph{w} (J/kg) is the specific work applied across the boundary \cite{1,2,3,4,StatThermo}
\begin{eqnarray}
\label{eq:eqWdef}
{\delta}w&=&{P}{\cdot}{{\delta}v}.
\end{eqnarray}
It is \emph{a priori} and intuitively obvious that energy cannot spontaneously appear from nowhere, and this principle is fundamental to thermodynamics.  

The second law has been described by Rudolph Clausius \cite{ClausiusOrig} in 1854 as \emph{heat can never pass from a colder to a warmer body without some other change, connected therewith, occurring at the same time.}  This principle is \emph{a posteriori} and consistently observed in nature that heat always flows from hot to cold.  The simple reason for this is the fact that due to kinetic theory \cite{2,3,KinTheoryBornGreen1946}, the square root of the temperature is proportional to average velocity of a particle $v_m$ (m/s)
\begin{eqnarray}
\label{eq:eqKT_VT}
v_m=\sqrt{\frac{{3}{\cdot}{\kappa}{\cdot}{T}}{m_m}},
\end{eqnarray}
where $\kappa$ represents the Boltzman's Constant (1.38$\mathrm{{\cdot}10^{-23}}$ J/K) and $m_m$ (kg) is the mass of a molecule.  When there is heat transfer, the higher velocity particle from the hotter matter transmits energy when it impacts the lower velocity molecule.  

In addition, Clausius' Theorem (\emph{a posteriori}) for the second law \cite{ClausiusOrig}
\begin{eqnarray}
\label{eq:eqClausius}
{{\oint}{\frac{{\delta}q}{T}}}&\leq&0,
\end{eqnarray}
states that any internally reversible thermodynamic cycle must generate a positive entropy ${\delta}s{\geq}0$ to the surrounding universe, where the change in entropy ${\delta}s$ (J/kg$\cdot$K) is defined as \cite{1,2,3,4,StatThermo}
\begin{eqnarray}
\label{eq:eqSideal}
{\delta}{s}&=&\frac{{\delta}q}{T},
\end{eqnarray}
where \emph{T} (K) is the absolute temperature, and ${\delta}q$ (J/kg) represent the heat transfered per unit mass.  

An internally reversible thermodynamic heat engine cycle with no increase in universal entropy ${\delta}s=0$ is the definition of the idealized Carnot efficiency $\eta_C$ of a heat engine \cite{ClausiusOrig,2}
\begin{eqnarray}
\label{eq:eqEfficC}
{\eta_C}={\frac{w_{out}}{q_{in}}}={\frac{{q_{in}}-{q_{out}}}{q_{in}}}={1-\frac{T_L}{T_H}},
\end{eqnarray}
where $w_{out}$ (J/kg) is the net work output, $q_{in}$ (J/kg) and $q_{out}$ (J/kg) are the heat input and output at the hot $T_H$ (K) and cold $T_L$ (K) temperatures, and $\eta_C$ represents the efficiency of a heat engine where there is no increase in entropy $\delta$\emph{s = 0},
\begin{eqnarray}
{\delta}s={\frac{q_{in}}{T_H}}-{\frac{q_{out}}{T_L}}=0,\nonumber \\ \nonumber
{\frac{q_{out}}{q_{in}}}={\frac{T_L}{T_H}},
\end{eqnarray}
and thus
\begin{eqnarray}
{\eta_C}={\frac{{q_{in}}-{q_{out}}}{q_{in}}}={1-{\frac{{q_{out}}}{q_{in}}}}={1-\frac{T_L}{T_H}}.\nonumber
\end{eqnarray}
A Carnot heat pump is simply a Carnot heat engine in reverse, and thus the Coefficient of Performance (COP) where $\delta$\emph{s=0} is
\begin{eqnarray}
\label{eq:eqCOPc}
{COP_C}={\frac{1}{\eta_C}}={\frac{q_{out}}{w_{in}}}={\frac{1}{1-\frac{T_L}{T_H}}}.
\end{eqnarray}
If a heat pump were designed so that the heat output would entirely supply the heat input of a heat engine, and then the work output of the heat engine would supply the work input of a heat pump, this system would run indefinitely provided that 
\begin{eqnarray}
\label{eq:eqPMM}
{\eta_{HE}}&\geq&{\frac{1}{COP_{HP}}},
\end{eqnarray}
and if equation \ref{eq:eqPMM} does not hold true, then a work input will be constantly needed to keep the heat-pump-heat-engine system running.  Since heat always flows from hot to cold, for this system to be possible the temperate range of the heat pump must be equal or greater than that of the heat engine
\begin{eqnarray}
{T_{H,HE}}&\leq&{T_{H,HP}},\nonumber \\ \nonumber {T_{L,HE}}&\geq&{T_{L,HP}},
\end{eqnarray}
and therefore if both the heat pump and heat engine maintained the ideal Carnot $COP_C$ and efficiency $\eta_C$, and the temperature difference was minimized so that $T_{H,HE} = T_{H,HP}$ and $T_{L,HE} = T_{L,HP}$, then $\eta_{HE} = {1}/{COP_{HP}}$.  If the heat pump or the heat engine ever exceeded the Carnot efficiency and $\eta_{HE}{>}{1}/{COP_{HP}}$, then Clausius' Theorem (equation \ref{eq:eqClausius}) would be violated and there would be negative entropy generated in the universe, and the system could obtain useful work from the ambient temperature, without the need for a temperature differential; this would violate Clausius' definition of the second law defined in equation \ref{eq:eqClausius}.  To date, no practical heat engine or heat pump that exceeded the Carnot efficiency has ever been demonstrated, though some examples have been demonstrated at the quantum level \cite{Carnot_PRL,Carnot_PRX,Carnot_NJOP}.  


The internal energy of an ideal gas is comprised solely of the kinetic energy and is \emph{only} affected by the temperature.  For a real gas, however, the intermolecular forces affect the behavior of the molecules \cite{1,2,3,4,Keesom2, KeesomOrig, LondonDispOrig, IntermolCermanic, RS_DispersionTemp, TempDepPhysRevB}.  The impacts of these forces increase as the molecules move closer together, and as the specific volume \emph{v} (m$^3$/kg) of the fluid decreases.  The current equation used to date for the change in specific internal energy \emph{u} (J/kg) for a real gas is based on the assumptions of entropy \cite{1,2} defined by Clausius in equation \ref{eq:eqClausius}
\begin{eqnarray}
\label{eq:eqdU_ideal}
{\delta}{u}&=&{{C_V}{\cdot}{{\delta}T}}+{\{{T{\cdot}{(\frac{{\partial}P}{{\partial}T}})}_V-{P}\}{\cdot}{{\delta}v}}, 
\end{eqnarray}
where $C_V$ (J/kg{$\cdot$}K) is the specific heat capacity at a constant volume.  The derivation of equation \ref{eq:eqdU_ideal} originates from the first law of thermodynamics defined in equation \ref{eq:eqFirstLaw}, which using equation \ref{eq:eqSideal}, the first law can then be written as
\begin{eqnarray}
\label{eq:eq_dUx01}
{{\delta}u}&=&{T{\cdot}{{\delta}s}}-{P{\cdot}{{\delta}v}}.
\end{eqnarray}
Expanding the partial derivatives of the entropy yields
\begin{eqnarray}
\label{eq:eq_dUx02}
{{\delta}s}&=&{{(\frac{{\partial}s}{{\partial}T})_V}{\cdot}{{\delta}T}}+{{(\frac{{\partial}s}{{\partial}V})_T}{\cdot}{{\delta}v}},
\end{eqnarray}
and due to the symmetry of the second derivative of the Helmholtz free energy \cite{2,3}
\begin{eqnarray}
\label{eq:eq_dUx03}
(\frac{{\partial}s}{{\partial}V})_T&=&(\frac{{\partial}P}{{\partial}T})_V.  
\end{eqnarray}
By plugging equation \ref{eq:eq_dUx03} into equation \ref {eq:eq_dUx02}, and then plugging equation \ref{eq:eq_dUx02} into equation \ref{eq:eq_dUx01}, and then defining the specific heat capacity
\begin{eqnarray}
{{T}{\cdot}{(\frac{{\partial}s}{{\partial}T})_V}{\cdot}{{\delta}T}}={{(\frac{q}{T})_V}{\cdot}{{\delta}T}}={{C_V}{\cdot}T},
\end{eqnarray}
one can get equation \ref{eq:eqdU_ideal}.  

This manuscript demonstrates a new thermodynamic cycle, utilizing a saturated fluid, where established and experimentally validated thermodynamic properties have been published by the \emph{National Institute of Standards and Technology} (NIST) \cite{NIST_Webbook}, formerly known as the \emph{National Bureau of Standards}.  On the assumption that the published thermodynamic properties, which have been used in research and industry for decades, are accurate, a theoretical thermodynamic cycle that generates a net negative entropy can be demonstrated, and thus disproving fundamental nature of Clausius' definition for the second law for dense fluids subjected to temperature-dependent attractive intermolecular forces.

\section{The Saturated Thermodynamic Cycle}

The theoretical cycle proposed starts off at a low temperature, saturated gas; this will be referred to as Stage 1.  A piston compresses the saturated gas isothermally until it is a saturated liquid (Stage 2), resulting in a decrease in internal energy, a mechanical work input, and a heat output to the cold-temperature sink.  Next, the piston expands slowly in a precise manner while the saturated fluid increases in temperature so that the fluid remains a saturated liquid until it is at a higher temperature; this hot saturated liquid will be referred to as Stage 3.  During this saturated liquid heating between stage 2 and 3, there is a heat input, an internal energy increase, and a (relatively minimal) mechanical work output.  Next, the piston continues to expand isothermally until the fluid is a saturated gas at the hot temperature; this will be referred to as Stage 4.  During the hot isothermal expansion, the internal energy will increase significantly, there will be a significant work output on the piston, and a significant heat input as well.  Finally, the piston will continue to expand precisely while the saturated gas is cooled, so that it remains a saturated gas, until the temperature returns to a saturated gas at the original cold temperature of Stage 1 and Stage 2.  During this saturated gas cooling, there is a decrease in internal energy, a work output, and (usually but not exclusively) a net heat input.

As the density of a fluid increases to the point of being a saturated liquid, saturated gas, or supercritical fluid, intermolecular attractive (and repulsive) forces \cite{KeesomOrig, Keesom2, LondonDispOrig, IntermolCermanic, RS_DispersionTemp, TempDepPhysRevB} can impact the pressure and temperature of the fluid.  As the molecules get closer together in the presence of attractive intermolecular forces, the internal potential energy will decrease.  
The thermodynamic data yields an empirical equation that closely predicts the change in specific internal energy ${{\Delta}u}$ (J/kg) during isothermal compression and expansion
\begin{eqnarray}
\label{eq:eqDeltaU_mytheory}
{{\Delta}u}&=&{{\frac{a'}{\sqrt{T}}}{\cdot}{({\frac{1}{v_1}}-{\frac{1}{v_2}})}},  \\ 
{a'}&=&\frac{{R^2}{\cdot}{T_c^{2.5}}}{9{\cdot}{({2^{\frac{1}{3}}-1})}{\cdot}{P_c}}.\nonumber 
\end{eqnarray}
where $v_1$ and $v_2$ (m$^3$/kg) represent the specific volume, \emph{T} represents the temperature, \emph{R} (J/kg${\cdot}$K) represents the gas constant, $T_C$ (K) represents the critical temperature, and $P_C$ (Pa) represents the critical pressure.  The value of \emph{a'} happens to be the same coefficient used in the Redlich-Kwong \cite{RK1949} equation of state; equation \ref{eq:eqDeltaU_mytheory} does not actually use any equation of state, as it is an empirical equation based on published data by NIST in the literature.  

For many different real fluids, this cycle will result in a net reduction in entropy ${\delta}s_u<0$ for the ambient universe, violating Clausius' Theorem defined in equation \ref{eq:eqClausius}.  In addition, due to Maxwell's Construction the process of a saturated liquid boiling into a saturated gas is a perfect example of constant-temperature expansion, and the change in internal energy for these ten different fluids boiling can be used to validate equation \ref{eq:eqDeltaU_mytheory} for the change in internal potential energy during isothermal expansion and compression.  




\subsection*{Propane}

This saturated thermodynamic cycle can contain any real fluid; this first example will contain propane ($\mathrm{{C_3}{H_8}}$) as the working fluid, utilizing the published thermodynamic properties \cite{nistCxHy} tabulated in Table \ref{tb:tbC3H8_NIST}.  This cycle will operate with the impractically large compression ratio of 983,286.  Propane has a molar mass of 44.098 grams per mole, which was used to tabulate the internal energy of a saturated liquid $u_L$ (J/kg) and a saturated gas $u_G$ (J/kg) in the last two columns of Table \ref{tb:tbC3H8_NIST}.  The saturated propane heat engine will operate at a low temperature of 130 K, and a hot temperature of 360 K.  

\begin{table}[h]
\begin{center}
\begin{tabular}{ | c || c | c | c | c | c |}
  \hline
$T_{sat}$ (K) & $P_{sat}$ (MPa) & ${{\rho}_L}$ (kg/m$^3$) & ${{\rho}_G}$ (kg/m$^3$) & $u_L$ (J/mole) & $u_L$ (J/kg)\\
  \hline
130 & 2e-5 & 688.3 & 0.0007 & -18060 & 3767\\
140 & 8e-5 & 678.1 & 0.003 & -17190 & 4149\\
150 & 2.8e-4 & 667.9 & 0.01 & -16310 & 4545\\
160 & 8.5e-4 & 657.7 & 0.0281 & -15420 & 4953\\
170 & 2.20e-3 & 647.4 & 0.0687 & -14530 & 5372\\
180 & 5.05e-3 & 637 & 0.1494 & -13630 & 5802\\
190 & .01051 & 626.5 & 0.2955 & -12710 & 6241\\
200 & .02013 & 615.8 & 0.5401 & -11780 & 6689\\
210 & .03593 & 605 & 0.9241 & -10840 & 7144\\
220 & .06044 & 593.9 & 1.496 & -9882 & 7606\\
230 & .09663 & 582.5 & 2.312 & -8903 & 8072\\
240 & .1479 & 570.7 & 3.435 & -7903 & 8543\\
250 & .2179 & 558.6 & 4.938 & -6882 & 9016\\
260 & .3107 & 546.1 & 6.905 & -5835 & 9490\\
270 & .4306 & 553 & 9.432 & -4762 & 9962\\
280 & .5819 & 519.2 & 12.64 & -3659 & 10430\\
290 & .7694 & 504.7 & 16.67 & -2523 & 10890\\
300 & .9978 & 478.3 & 21.7 & -1352 & 11340\\
310 & 1.272 & 472.6 & 27.99 & -139.6 & 11760\\
320 & 1.598 & 454.5 & 35.89 & 1121 & 12160\\
330 & 1.982 & 434.4 & 45.93 & 2440 & 12510\\
340 & 2.431 & 411.3 & 59.01 & 3836 & 12790\\
350 & 2.954 & 383.4 & 76.99 & 5347 & 12950\\
360 & 3.564 & 345.6 & 105 & 7073 & 12850\\
  \hline
\end{tabular}
\caption{Saturated thermodynamic properties for propane $\mathrm{{C_3}{H_8}}$ from the \emph{National Institute of Standards and Technology} \cite{nistCxHy}.}
\label{tb:tbC3H8_NIST}
\end{center}
\end{table}

First, the published thermodynamic properties \cite{nistCxHy} for propane tabulated in Table \ref{tb:tbC3H8_NIST} can also be used to validate equation \ref{eq:eqDeltaU_mytheory}.  Propane has a critical temperature $T_C$ of 369.85 K, a critical pressure $P_C$ of 4.24766 MPa, a critical density of 220.5 kg/m$\mathrm{^3}$, and a molar mass of 44.098 g/mole; the coefficient \emph{a'} defined in equation \ref{eq:eqDeltaU_mytheory} is therefore 9,410.4979 {Pa}${\cdot}${K$^{0.5}$}$\cdot${{m$^6$}${\cdot}$kg$^{-2}$}.  The difference in the experimentally validated change in internal energy during vaporization, as well as the calculated change in internal energy for isothermal expansion using equation \ref{eq:eqDeltaU_mytheory} is tabulated in Table \ref{tb:tbC3H8_dU}.  The average error for propane is 8.8375\%, and the coefficient of determination $R^2$ value is 0.96625, validating equation \ref{eq:eqDeltaU_mytheory}.  
\begin{table}[h]
\begin{center}
\begin{tabular}{ | c || c | c | c | c | c |}
  \hline
T (K) & $u_L$ (J/kg) & $u_L$ (J/kg) & $\delta{u}$ (J/kg) exp & $\delta{u}$ (J/kg) calc & Error (\%)\\
  \hline
130 & -409,542.4 & 85,423.4 & 494,966 & 568,091 & 14.77\%\\
140 & -389,813.6 & 94,085.9 & 483,899 & 539,312 & 11.45\%\\
150 & -369,858.0 & 103,065.9 & 472,924 & 513,183 & 8.51\%\\
160 & -349,675.7 & 112,318.0 & 461,994 & 489,285 & 5.91\%\\
170 & -329,493.4 & 121,819.6 & 451,313 & 467,213 & 3.52\%\\
180 & -309,084.3 & 131,570.6 & 440,655 & 446,698 & 1.37\%\\
190 & -288,221.7 & 141,525.7 & 429,747 & 427,516 & 0.52\%\\
200 & -267,132.3 & 151,684.9 & 418,817 & 409,408 & 2.25\%\\
210 & -245,816.1 & 162,002.8 & 407,819 & 392,278 & 3.81\%\\
220 & -224,091.8 & 172,479.5 & 396,571 & 375,854 & 5.22\%\\
230 & -201,891.2 & 183,046.9 & 384,938 & 360,012 & 6.48\%\\
240 & -179,214.5 & 193,727.6 & 372,942 & 344,582 & 7.60\%\\
250 & -156,061.5 & 204,453.7 & 360,515 & 329,524 & 8.60\%\\
260 & -132,318.9 & 215,202.5 & 347,521 & 314,682 & 9.45\%\\
270 & -107,986.8 & 225,905.9 & 333,893 & 299,850 & 10.20\%\\
280 & -82,974.3 & 236,518.7 & 319,493 & 284,882 & 10.83\%\\
290 & -57,213.5 & 246,950.0 & 304,163 & 269,687 & 11.33\%\\
300 & -30,659.0 & 257,154.5 & 287,814 & 248,078 & 13.81\%\\
310 & -3,165.7 & 266,678.8 & 269,844 & 237,635 & 11.94\%\\
320 & 25,420.7 & 275,749.5 & 250,329 & 220,215 & 12.03\%\\
330 & 55,331.3 & 283,686.3 & 228,355 & 201,239 & 11.87\%\\
340 & 86,988.1 & 290,035.8 & 203,048 & 179,793 & 11.45\%\\
350 & 121,252.7 & 293,664.1 & 172,411 & 154,128 & 10.60\%\\
360 & 160,392.8 & 291,396.4 & 131,004 & 119,332 & 8.91\%\\
   \hline
\end{tabular}
\caption{The change in internal energy during the vaporization of propane.}
\label{tb:tbC3H8_dU}
\end{center}
\end{table}

The change in internal energy during the isothermal compression ${\delta}{u_{12}}$ (J/kg) can be easily determined with the first row of Table \ref{tb:tbC3H8_NIST}
\begin{eqnarray}
\label{eq:eqdU12}
{\delta}{u_{12}}={u_2}-{u_1}=-409,542.5-85,423.5=-494,966. 
\end{eqnarray}
Due to Maxwell's Construction, the pressure during condensation remains constant, and therefore the work input ${w_{12}}$ (J/kg) defined in equation \ref{eq:eqWdef} for the isothermal compression is simply
\begin{eqnarray}
\label{eq:eqW12}
{w_{12}}={P_s}{\cdot}({\frac{1}{\rho_L}}-{\frac{1}{\rho_G}})={20}{\cdot}({\frac{1}{0.0007}}-{\frac{1}{688.3}})=28,571. 
\end{eqnarray}
Utilizing the first law of thermodynamics (equation \ref{eq:eqFirstLaw}), the total heat output ${q_{12}}$ (J/kg) during isothermal compression is therefore
\begin{eqnarray}
\label{eq:eqQ12}
{q_{12}}={{\delta}{u_{12}}}-{w_{12}}={-494,966}-{28,571}=-523,537. 
\end{eqnarray}
The change in entropy of the ambient universe during isothermal compression ${\delta}{s_{12}}$ (J/kg$\cdot$K), defined in equation \ref{eq:eqSideal}, can thus be determined
\begin{eqnarray}
\label{eq:eqS12}
{\delta}{s_{12}}&=&{-\frac{{q_{12}}}{T_{12}}}={\frac{523,537}{130}}=4,027.2.
\end{eqnarray}

The experimentally verified thermodynamic properties tabulated in Table \ref{tb:tbC3H8_NIST} were used to generated Table \ref{tb:tbC3H8_CalcL}, where the saturated liquid is heated, with expansion in order that the propane remains a saturated liquid as the temperature increases.  The piston expands minimally (${\delta}{v_{23}}$=1,440 cm$^3$/kg), and the work output (${w_{23}}$=-2,227 J/kg) is small compared to the change in internal energy (${{\delta}u_{23}}$=569,935 J/kg), and thus the total heat input $q_{23}$ (J/kg) during the heating of the saturated liquid propane is thus 
\begin{eqnarray}
\label{eq:eqQ23}
{q_{23}}={{\delta}u_{23}}-{w_{23}}={569,935}-{(-2,227)}=572,162.
\end{eqnarray}
The total reduction in global entropy during this heating is ${\delta}s_{23}$=-2,408.43 J/kg$\cdot$K.  

\begin{table}[h]
\begin{center}
\begin{tabular}{ | c || c | c | c | c | c | c |}
  \hline
$T_{avg}$ (K) & $P_{avg}$ (Pa) & ${{\delta}{v}_L}$ (cm$^3$/kg) & ${\delta}u_L$ (J/kg) & $w_L$ (J/kg) & $q_L$ (J/kg) & ${\delta}s_L$ (J/kg$\cdot$K)\\
  \hline
135 & 50 & 21.85 & 19,729 & 0.0011 & 19,729 & -146.14\\
145 & 180 & 22.52 & 19,956 & 0.0041 & 19,956 & -137.62\\
155 & 565 & 23.22 & 20,182 & 0.0131 & 20,182 & -130.21\\
165 & 1,525 & 24.19 & 20,182 & 0.0369 & 20,182 & -122.32\\
175 & 3,625 & 25.22 & 20,409 & 0.0914 & 20,409 & -116.62\\
185 & 7,780 & 26.31 & 20,863 & 0.2047 & 20,863 & -112.77\\
195 & 15,320 & 27.73 & 21,089 & 0.4249 & 21,090 & -108.15\\
205 & 28,030 & 28.99 & 21,316 & 0.8126 & 21,317 & -103.99\\
215 & 48,185 & 30.89 & 21,724 & 1.49 & 21,726 & -101.05\\
225 & 78,535 & 32.95 & 22,201 & 2.59 & 22,203 & -98.68\\
235 & 122,265 & 35.50 & 22,677 & 4.34 & 22,681 & -96.52\\
245 & 182,900 & 37.96 & 23,153 & 6.94 & 23,160 & -94.53\\
255 & 264,300 & 40.98 & 23,743 & 10.83 & 23,753 & -93.15\\
265 & 370,650 & 45.01 & 24,332 & 16.68 & 24,349 & -91.88\\
275 & 506,250 & 49.87 & 25,012 & 25.25 & 25,038 & -91.05\\
285 & 675,650 & 55.34 & 25,761 & 37.39 & 25,798 & -90.52\\
295 & 883,600 & 109.36 & 26,554 & 96.63 & 26,651 & -90.34\\
305 & 1,134,900 & 25.22 & 27,493 & 28.62 & 27,522 & -90.24\\
315 & 1,435,000 & 84.27 & 28,586 & 120.92 & 28,707 & -91.13\\
325 & 1,790,000 & 101.81 & 29,911 & 182.23 & 30,093 & -92.59\\
335 & 2,206,500 & 129.29 & 31,657 & 285.28 & 31,942 & -95.35\\
345 & 2,692,500 & 176.93 & 34,265 & 476.38 & 34,741 & -100.70\\
355 & 3,259,000 & 285.28 & 39,140 & 929.72 & 40,070 & -112.87\\
  \hline
Total &  & 1,440.66 & 569,935 & 2,227 & 572,162 & -2,408.43\\
  \hline
\end{tabular}
\caption{Thermodynamic properties during the heating of saturated liquid propane.}
\label{tb:tbC3H8_CalcL}
\end{center}
\end{table}

The change in internal energy during the isothermal expansion ${\delta}{u_{34}}$ (J/kg) can be easily determined with the first row of Table \ref{tb:tbC3H8_NIST}
\begin{eqnarray}
\label{eq:eqU34}
{\delta}{u_{34}}={u_4}-{u_3}=291,397-160,393=131,004.
\label{eq:eqdU34}
\end{eqnarray}
The pressure during isothermal vaporization remains constant, and therefore the work output ${w_{34}}$ (J/kg) defined in equation \ref{eq:eqWdef} for the isothermal expansion is simply
\begin{eqnarray}
\label{eq:eqW34}
{w_{34}}={P_s}{\cdot}({\frac{1}{\rho_L}}-{\frac{1}{\rho_G}})={3,564,000}{\cdot}({\frac{1}{345.6}}-{\frac{1}{105}})=-23,631. 
\end{eqnarray}
Utilizing the first law of thermodynamics (equation \ref{eq:eqFirstLaw}), the total heat output ${q_{34}}$ (J/kg) during isothermal compression is therefore
\begin{eqnarray}
\label{eq:eqQ34}
{q_{34}}={{\delta}{u_{34}}}-{w_{34}}={131,004}-{(-23,631)}=154,635. 
\end{eqnarray}
The change in entropy of the ambient universe during isothermal compression ${\delta}{s_{34}}$ (J/kg$\cdot$K), defined in equation \ref{eq:eqSideal}, can thus be determined
\begin{eqnarray}
\label{eq:eqS34}
{\delta}{s_{34}}&=&{-\frac{{q_{34}}}{T_{34}}}={\frac{154,635}{360}}=-429.54.
\end{eqnarray}

Finally, the experimentally verified thermodynamic properties tabulated in Table \ref{tb:tbC3H8_NIST} were used to generated Table \ref{tb:tbC3H8_CalcG}, where the saturated gas is cooled, with expansion in order that the propane remain a saturated gas as the temperature decreases.  As the saturated propane gas is cooling, the internal energy is decreasing, where ${{\delta}u_{41}}$=-205,973 J/kg.  While the heating of the saturated liquid resulted in a very small increase in specific volume of ${\delta}{v_{23}}$=1,440 cm$^3$/kg, the cooling of the saturated gas results in a dramatic increase in specific volume of ${\delta}{v_{41}}$=1,428.56 m$^3$/kg, nearly six orders of magnitude greater expansion.  As a result of this large expansion, there is a significant work output of ${w_{41}}$=-471,475 J/kg, and therefore despite the cooling and decrease in internal energy, there is a net heat input ${q_{41}}$ (J/kg) during saturated gas expansion
\begin{eqnarray}
\label{eq:eqQ41}
{q_{41}}={{\delta}u_{41}}-{w_{41}}={-205,973}-{(-471,475)}=265,502.
\end{eqnarray}
Because of this heat input, there is a decrease in global entropy during the saturated gas cooling of ${\delta}s_{41}$=-1,479.76 J/kg$\cdot$K.  

\begin{table}[h]
\begin{center}
\begin{tabular}{ | c || c | c | c | c | c | c |}
  \hline
$T_{avg}$ (K) & $P_{avg}$ (Pa) & ${{\delta}{v}_G}$ (m$^3$/kg) & ${\delta}u_G$ (J/kg) & $w_G$ (J/kg) & $q_G$ (J/kg) & ${\delta}s_G$ (J/kg$\cdot$K)\\
  \hline
135 & 50 & 1,095.238 & -8,663 & 54,762 & 46,099 & -341.48\\
145 & 180 & 233.333 & -8,980 & 42,000 & 33,020 & -227.72\\
155 & 565 & 64.413 & -9,252 & 36,393 & 27,141 & -175.10\\
165 & 1,525 & 21.031 & -9,502 & 32,073 & 22,571 & -136.79\\
175 & 3,625 & 7.863 & -9,751 & 28,502 & 18,751 & -107.15\\
185 & 7,780 & 3.309 & -9,955 & 25,747 & 15,792 & -85.36\\
195 & 15,320 & 1.533 & -10,159 & 23,479 & 13,320 & -68.31\\
205 & 28,030 & 0.769 & -10,318 & 21,566 & 11,248 & -54.87\\
215 & 48,185 & 0.414 & -10,477 & 19,933 & 9,457 & -43.98\\
225 & 78,535 & 0.236 & -10,567 & 18,528 & 7,961 & -35.38\\
235 & 122,265 & 0.141 & -10,681 & 17,289 & 6,608 & -28.12\\
245 & 182,900 & 0.0886 & -10,726 & 16,207 & 5,481 & -22.37\\
255 & 264,300 & 0.0577 & -10,749 & 15,247 & 4,498 & -17.64\\
265 & 370,650 & 0.0388 & -10,703 & 14,381 & 3,678 & -13.88\\
275 & 506,250 & 0.0269 & -10,613 & 13,622 & 3,010 & -10.94\\
285 & 675,650 & 0.0191 & -10,431 & 12,922 & 2,491 & -8.74\\
295 & 883,600 & 0.0139 & -10,205 & 12,287 & 2,082 & -7.06\\
305 & 1,134,900 & 0.0104 & -9,524 & 11,753 & 2,229 & -7.31\\
315 & 1,435,000 & 0.0079 & -9,071 & 11,285 & 2,214 & -7.03\\
325 & 1,790,000 & 0.0061 & -7,937 & 10,902 & 2,965 & -9.12\\
335 & 2,206,500 & 0.0048 & -6,349 & 10,649 & 4,299 & -12.83\\
345 & 2,692,500 & 0.0040 & -3,628 & 10,656 & 7,028 & -20.37\\
355 & 3,259,000 & 0.0035 & 2,268 & 11,292 & 13,560 & -38.20\\
  \hline
Total &  & 1,428.56 & -205,973 & 471,475 & 265,502 & -1,479.76\\
  \hline
\end{tabular}
\caption{Thermodynamic properties during the cooling of saturated propane vapor.}
\label{tb:tbC3H8_CalcG}
\end{center}
\end{table}

In order for this cycle to be internally reversible, the net energy into and out of this heat engine cycle must balance.  This can be easily verified by taking the cumulation of the heat transfer and work during each of the four stages.
\begin{eqnarray}
\label{eq:eqQWnet}
{q_{12}}+{q_{23}}+{q_{34}}+{q_{41}}+{w_{12}}+{w_{23}}+{w_{34}}+{w_{41}}&=&0\nonumber \\ \nonumber
{(-523,537+572,162+154,635+265,502)}+{(28,571-2,227-23,631-471,475)}&=&0
\end{eqnarray}
What is most remarkable about this cycle is the fact that the net entropy to the ambient universe is negative!
\begin{eqnarray}
\label{eq:eqDSu}
{{\delta}s_u}&=&{{\delta}s_{12}}+{{\delta}s_{23}}+{{\delta}s_{34}}+{{\delta}s_{41}}\nonumber \\
&=&{(4,027.2)}+{(-2,408.43)}+{(-429.54)}+{(-1,479.76)}\nonumber \\
&=&-290.53,\nonumber 
\end{eqnarray}
which contradicts Clausius Theorem for the second law of thermodynamics (equation \ref{eq:eqClausius}), all the while never allowing impossible heat transfer from cold to hot!


\section{Parametric Study}

The cycle was studied again parametrically for different fluids, both the highly polar fluid water; the monatomic fluids of argon, krypton, and xenon; the diatomic fluid nitrogen; ammonia; the hydrocarbons of methane, ethane, propane, and both normal and iso-butane; and the refrigerants Freon R-12, R-22, and R-134a.  All of the data provided utilized the available online tables from NIST \cite{NIST_Webbook}, which are based on previously published experimental and empirical thermodynamics data \cite{New_NIST_R134a,New_NIST_N2,New_NIST_H2O,New_NIST_CH4, New_NIST_C2H6, New_NIST_C3H8, New_NIST_C4H10n, New_NIST_C4H10iso, NISTargon1, NISTargon2, ArgonCV, ArgonHvThesis, ArgonCriticalProp, nistXn, Beattie_1951_Xn, HvNobelGases, nistN2, nistAmmonia, NISTsteamHvEqu, NISTwaterVolDat1, NISTwaterCritProp, NISTwaterVolDat2, NISTsteamDataTable, GoffGratch1946,nistCxHy,BWR1940}.  First, equation \ref{eq:eqDeltaU_mytheory} matched remarkably for the change in internal energy during isothermal expansion during vaporization, all over a wide temperature range ${\Delta}T$ (K).  The calculated coefficient \emph{a'} ({Pa}${\cdot}${K$^{0.5}$}$\cdot${{m$^6$}${\cdot}$kg$^{-2}$}) and the coefficient of determination $R^2$ between the NIST values and equation \ref{eq:eqDeltaU_mytheory} are all tabulated in Table \ref{tb:tbParamDU}.

\begin{table}[h]
\begin{center}
\begin{tabular}{ | c || c | c | c | c | c | c |}
  \hline
Fluid & M (g/Mole) & $T_C$ (K) & $P_C$ (MPa) & {a'} & ${\Delta}T$ (K) & $R^2$\\
  \hline
Water ($\mathrm{H_2O}$) & 18.02 & 647.14 & 22.064 & 43,971 & 274-647 & 0.98572\\
Argon (Ar) & 39.948 & 150.687 & 4.863 & 1,062 & 84-150 & 0.98911\\
Krypton (Kr) & 83.798 & 209.48 & 5.525 & 484 & 116-209 & 0.98858\\
Xenon (Xe) & 131.3 & 289 & 5.84 & 417 & 162-289 & 0.98972\\
Nitrogen ($\mathrm{N_2}$) & 28.0134 & 126.2 & 3.4 & 1,982 & 64-126 & 0.98565\\
Ammonia ($\mathrm{NH_3}$) & 17.0305 & 405.4 & 11.3119 & 29,824 & 196-405 & 0.98603\\
Methane ($\mathrm{CH_4}$) & 16.043 & 190.53 & 4.598 & 12,520 & 91-190 & 0.97818\\
Ethane ($\mathrm{C_2H_6)}$ & 30.07 & 305.34 & 4.8714 & 10,937 & 91-305 & 0.94881\\
Propane ($\mathrm{C_3H_8)}$ & 44.098 & 369.85 & 4.2477 & 9,418 & 86-369 & 0.93372\\
Butane ($\mathrm{C_4H_{10}}$) & 58.125 & 425.16 & 3.796 & 8,594 & 135-424 & 0.9631\\
Iso-Butane ($\mathrm{C_4H_{10}}$) & 58.125 & 407.85 & 3.64 & 8,078 & 114-407 & 0.95368\\
Freon R-12 & 120.91 & 385.12 & 4.1361 & 1,423 & 175-384 & 0.98465\\
Freon R-22 & 86.47 & 369.295 & 4.99 & 2,077 & 172-369 & 0.98741\\
Freon R-134a & 102.03 & 374.21 & 4.0593 & 1,896 & 170-374 & 0.9884\\
  \hline
\end{tabular}
\caption{The calculated coefficient \emph{a'} ({Pa}${\cdot}${K$^{0.5}$}$\cdot${{m$^6$}${\cdot}$kg$^{-2}$}) and the coefficient of determination $R^2$ between the NIST values (and equation \ref{eq:eqdU_ideal}) and equation \ref{eq:eqDeltaU_mytheory}, over a specified temperature range ${\Delta}T$ (K). }
\label{tb:tbParamDU}
\end{center}
\end{table}

It shall be noted that the change in internal energy ${\delta}u$ (J/kg) obtained from the NIST database \cite{NIST_Webbook} matches perfectly $R^2=1$ for the change in internal energy for isothermal (\emph{dT=0}) compression and expansion defined in equation \ref{eq:eqdU_ideal} 
\begin{eqnarray}
{\delta}{u}_{dT=0}&=&{\{{T{\cdot}{(\frac{{\partial}P}{{\partial}T}})}_V-{P}\}{\cdot}{{\delta}v}}, \nonumber
\end{eqnarray}
when using the tabulated saturated pressures and saturated specific volumes for ${({\frac{{\partial}P}{{\partial}T}})}_V$, \emph{P}, and ${{\delta}v}={{v_g}-{v_l}}$.  This is expected, as all of the equations of state used to generate the NIST table are based on equation \ref{eq:eqdU_ideal}; the close match between equation \ref{eq:eqdU_ideal} and equation \ref{eq:eqDeltaU_mytheory} offers further evidence of the validity of equation \ref{eq:eqDeltaU_mytheory} when calculating the change in internal energy of a real fluid during isothermal compression and expansion.  

Next, the saturated thermodynamic cycle was studied for all of these fluids.  While all of them had a ridiculously large compression ratio $\Phi$ making the cycle as described impractical to build, all of the fluids managed to reduce the net global entropy ${\delta}s_u$ (J/kg$\cdot$K) when implemented over a specified temperature range ${\Delta}T$; this reduction in net-global entropy is tabulated in Table \ref{tb:tbParamDS}.  The specific work input and output \emph{w}, heat input and output \emph{q}, change in internal energy \emph{u}, and change in entropy of the universe $\delta_s$ of this cycle is tabulated for water and steam (Table \ref{tb:tbWUQS_h2o}), argon (Table \ref{tb:tbWUQS_Ar}), krypton (Table \ref{tb:tbWUQS_Kr}), xenon (Table \ref{tb:tbWUQS_Xe}), nitrogen (Table \ref{tb:tbWUQS_n2}), ammonia (Table \ref{tb:tbWUQS_nh3}), methane (Table \ref{tb:tbWUQS_ch4}), ethane (Table \ref{tb:tbWUQS_c2h6}), propane (Table \ref{tb:tbWUQS_c3h8}), normal butane (Table \ref{tb:tbWUQS_c4h10n}), iso-butane (Table \ref{tb:tbWUQS_c4h10iso}), Freon R12 (Table \ref{tb:tbWUQS_r12}), Freon R22 (Table \ref{tb:tbWUQS_r22}), and Freon R134a (Table \ref{tb:tbWUQS_r134a}).  By utilizing experimentally validated thermodynamic values from NIST, it is clear that Clausius' interpretation of the second law defined in equation \ref{eq:eqClausius} is not universally applicable in the case of real fluids with temperature-dependent intermolecular attractive forces.  

\begin{table}[h]
\begin{center}
\begin{tabular}{ | c || c | c | c |}
  \hline
Fluid & ${\Delta}T$ (K) & {$\Phi$} & ${\delta}s_u$\\
  \hline
Water ($\mathrm{H_2O}$) & 274-624 & 194,392 & -635.48\\
Argon (Ar) & 84-111 & 343 & -76.25\\
Krypton (Kr) & 116-154 & 366 & -35.44\\
Xenon (Xe) & 162-212 & 349 & -21.2\\
Nitrogen ($\mathrm{N_2}$) & 64-103 & 1,112 & -435.76\\
Ammonia ($\mathrm{NH_3}$) & 196-324 & 11,004 & -961.83\\
Methane ($\mathrm{CH_4}$) & 91-161 & 1,736 & -1,184.07\\
Ethane ($\mathrm{C_2H_6}$) & 91-277 & 12,217,233 & -1,882.29\\
Propane ($\mathrm{C_3H_8}$) & 86-349 & 56,413,242,511 & -2,561.69\\
Butane ($\mathrm{C_4H_{10}}$) & 135-414 & 20,898,002 & -2,636.24\\
Iso-Butane ($\mathrm{C_4H_{10}}$) & 114-397 & 491,629,663 & -2,612.54\\
Freon R-12 & 175-364 & 14,264 & -549.14\\
Freon R-22 & 172-336 & 14,405 & -461.02\\
Freon R-134a & 170-355 & 55,572 & -743.06\\
  \hline
\end{tabular}
\caption{Reduction in global entropy ${\delta}s_u$ (J/kg$\cdot$K) from saturated thermodynamic cycle for various fluids, temperature ranges ${\Delta}T$ (K), and compression ratios $\Phi$.}
\label{tb:tbParamDS}
\end{center}
\end{table}



\clearpage


\begin{table}[h]
\begin{center}
\begin{tabular}{ | c || c | c | c | c |}
  \hline
Stage & 12 & 23 & 34 & 41\\
  \hline
\emph{w} & 126383 & -5803 & -115104 & -1672652\\
${\delta}u$ & -2372556 & 1964256 & 764800 & 304500\\
\emph{q} & -2498939 & 1650959 & 879904 & 1710052\\
${\delta}s$ & 9120.22 & -3776.21 & -1410.1 & -4569.39\\
  \hline
\end{tabular}
\caption{Work \emph{w} (J/kg) input and output, change in internal energy ${\delta}{u}$ \emph{W} (J/kg), heat \emph{u} (J/kg) input and output, and change in entropy ${\delta}s$ (J/kg$\cdot$K) for each stage of the cycle using Water/Steam (H$\mathrm{_2O}$) as the working fluid.}
\label{tb:tbWUQS_h2o}
\end{center}
\end{table}

\begin{table}[h]
\begin{center}
\begin{tabular}{ | c || c | c | c | c |}
  \hline
Stage & 12 & 23 & 34 & 41\\
  \hline
\emph{w} & 16973 & -33 & -19459 & -40122\\
${\delta}u$ & -146605 & 96432 & 119329 & 25815\\
\emph{q} & -163578 & 30832 & 138788 & 43645\\
${\delta}s$ & 1947.36 & -317.72 & -1250.34 & -455.55\\
  \hline
\end{tabular}
\caption{Work \emph{w} (J/kg) input and output, change in internal energy ${\delta}{u}$ \emph{W} (J/kg), heat \emph{u} (J/kg) input and output, and change in entropy ${\delta}s$ (J/kg$\cdot$K) for each stage of the cycle using Argon (Ar) as the working fluid.}
\label{tb:tbWUQS_Ar}
\end{center}
\end{table}

\begin{table}[h]
\begin{center}
\begin{tabular}{ | c || c | c | c | c |}
  \hline
Stage & 12 & 23 & 34 & 41\\
  \hline
\emph{w} & 11173 & -20 & -12860 & -26995\\
${\delta}u$ & -97150 & 66437 & 79060 & 19388\\
\emph{q} & -108323 & 20382 & 91920 & 29267\\
${\delta}s$ & 933.82 & -151.65 & -596.88 & -220.73\\
  \hline
\end{tabular}
\caption{Work \emph{w} (J/kg) input and output, change in internal energy ${\delta}{u}$ \emph{W} (J/kg), heat \emph{u} (J/kg) input and output, and change in entropy ${\delta}s$ (J/kg$\cdot$K) for each stage of the cycle using Krypton (Kr) as the working fluid.}
\label{tb:tbWUQS_Kr}
\end{center}
\end{table}

\begin{table}[h]
\begin{center}
\begin{tabular}{ | c || c | c | c | c |}
  \hline
Stage & 12 & 23 & 34 & 41\\
  \hline
\emph{w} & 9946 & -14 & -11378 & -22803\\
${\delta}u$ & -86328 & 58246 & 70822 & 16972\\
\emph{q} & -96274 & 17412 & 82200 & 24695\\
${\delta}s$ & 594.28 & -93.51 & -387.74 & -134.23\\
  \hline
\end{tabular}
\caption{Work \emph{w} (J/kg) input and output, change in internal energy ${\delta}{u}$ \emph{W} (J/kg), heat \emph{u} (J/kg) input and output, and change in entropy ${\delta}s$ (J/kg$\cdot$K) for each stage of the cycle using Xenon (Xe) as the working fluid.}
\label{tb:tbWUQS_Xe}
\end{center}
\end{table}

\begin{table}[h]
\begin{center}
\begin{tabular}{ | c || c | c | c | c |}
  \hline
Stage & 12 & 23 & 34 & 41\\
  \hline
\emph{w} & 18779 & -120 & -22822 & -86886\\
${\delta}u$ & -195844 & 157583 & 131103 & 17321\\
\emph{q} & -214623 & 81622 & 153925 & 103647\\
${\delta}s$ & 3353.48 & -990.88 & -1494.42 & -1303.94\\
  \hline
\end{tabular}
\caption{Work \emph{w} (J/kg) input and output, change in internal energy ${\delta}{u}$ \emph{W} (J/kg), heat \emph{u} (J/kg) input and output, and change in entropy ${\delta}s$ (J/kg$\cdot$K) for each stage of the cycle using Nitrogen (N$\mathrm{_2}$) as the working fluid.}
\label{tb:tbWUQS_n2}
\end{center}
\end{table}

\begin{table}[h]
\begin{center}
\begin{tabular}{ | c || c | c | c | c |}
  \hline
Stage & 12 & 23 & 34 & 41\\
  \hline
\emph{w} & 95263 & -274 & -125084 & -629096\\
${\delta}u$ & -1387977 & 1151977 & 921020 & 125900\\
\emph{q} & -1483240 & 582531 & 1046104 & 744396\\
${\delta}s$ & 7567.55 & -2272.82 & -3228.72 & -3027.84\\
  \hline
\end{tabular}
\caption{Work \emph{w} (J/kg) input and output, change in internal energy ${\delta}{u}$ \emph{W} (J/kg), heat \emph{u} (J/kg) input and output, and change in entropy ${\delta}s$ (J/kg$\cdot$K) for each stage of the cycle using Ammonia (NH$\mathrm{_3}$) as the working fluid.}
\label{tb:tbWUQS_nh3}
\end{center}
\end{table}

\begin{table}[h]
\begin{center}
\begin{tabular}{ | c || c | c | c | c |}
  \hline
Stage & 12 & 23 & 34 & 41\\
  \hline
\emph{w} & 46766 & -487 & -57568 & -263440\\
${\delta}u$ & -497066 & 426576 & 310120 & 3320\\
\emph{q} & -543832 & 255553 & 367688 & 331560\\
${\delta}s$ & 5976.18 & -2062.67 & -2283.78 & -2813.8\\
  \hline
\end{tabular}
\caption{Work \emph{w} (J/kg) input and output, change in internal energy ${\delta}{u}$ \emph{W} (J/kg), heat \emph{u} (J/kg) input and output, and change in entropy ${\delta}s$ (J/kg$\cdot$K) for each stage of the cycle using Methane (CH$\mathrm{_4}$) as the working fluid.}
\label{tb:tbWUQS_ch4}
\end{center}
\end{table}

\begin{table}[h]
\begin{center}
\begin{tabular}{ | c || c | c | c | c |}
  \hline
Stage & 12 & 23 & 34 & 41\\
  \hline
\emph{w} & 25161 & -880 & -43862 & -537857\\
${\delta}u$ & -570420 & 611670 & 242780 & -88460\\
\emph{q} & -595581 & 474980 & 286642 & 684317\\
${\delta}s$ & 6544.85 & -2765.78 & -1034.81 & -4626.55\\
  \hline
\end{tabular}
\caption{Work \emph{w} (J/kg) input and output, change in internal energy ${\delta}{u}$ \emph{W} (J/kg), heat \emph{u} (J/kg) input and output, and change in entropy ${\delta}s$ (J/kg$\cdot$K) for each stage of the cycle using Ethane (C$\mathrm{_2H_6}$) as the working fluid.}
\label{tb:tbWUQS_c2h6}
\end{center}
\end{table}

\begin{table}[h]
\begin{center}
\begin{tabular}{ | c || c | c | c | c |}
  \hline
Stage & 12 & 23 & 34 & 41\\
  \hline
\emph{w} & 16218 & -1239 & -31176 & -628559\\
${\delta}u$ & -545100 & 705840 & 176050 & -213350\\
\emph{q} & -561318 & 612119 & 207226 & 870389\\
${\delta}s$ & 6526.95 & -3074.67 & -593.77 & -5420.2\\
  \hline
\end{tabular}
\caption{Work \emph{w} (J/kg) input and output, change in internal energy ${\delta}{u}$ \emph{W} (J/kg), heat \emph{u} (J/kg) input and output, and change in entropy ${\delta}s$ (J/kg$\cdot$K) for each stage of the cycle using Propane (C$\mathrm{_3H_8}$) as the working fluid.}
\label{tb:tbWUQS_c3h8}
\end{center}
\end{table}

\begin{table}[h]
\begin{center}
\begin{tabular}{ | c || c | c | c | c |}
  \hline
Stage & 12 & 23 & 34 & 41\\
  \hline
\emph{w} & 19312 & -1895 & -20572 & -438105\\
${\delta}u$ & -476527 & 740977 & 120460 & -314540\\
\emph{q} & -495839 & 688622 & 141032 & 768765\\
${\delta}s$ & 3672.88 & -2622.42 & -340.66 & -3346.04\\
  \hline
\end{tabular}
\caption{Work \emph{w} (J/kg) input and output, change in internal energy ${\delta}{u}$ \emph{W} (J/kg), heat \emph{u} (J/kg) input and output, and change in entropy ${\delta}s$ (J/kg$\cdot$K) for each stage of the cycle using Normal Butane (C$\mathrm{_4H_{10}}$) as the working fluid.}
\label{tb:tbWUQS_c4h10n}
\end{center}
\end{table}

\begin{table}[h]
\begin{center}
\begin{tabular}{ | c || c | c | c | c |}
  \hline
Stage & 12 & 23 & 34 & 41\\
  \hline
\emph{w} & 16307 & -1853 & -19879 & -472977\\
${\delta}u$ & -464170 & 707960 & 113580 & -288420\\
\emph{q} & -480477 & 657283 & 133459 & 777817\\
${\delta}s$ & 4214.71 & -2703.68 & -336.17 & -3787.4\\
  \hline
\end{tabular}
\caption{Work \emph{w} (J/kg) input and output, change in internal energy ${\delta}{u}$ \emph{W} (J/kg), heat \emph{u} (J/kg) input and output, and change in entropy ${\delta}s$ (J/kg$\cdot$K) for each stage of the cycle using Iso-Butane (C$\mathrm{_4H_{10}}$) as the working fluid.}
\label{tb:tbWUQS_c4h10iso}
\end{center}
\end{table}

\begin{table}[h]
\begin{center}
\begin{tabular}{ | c || c | c | c | c |}
  \hline
Stage & 12 & 23 & 34 & 41\\
  \hline
\emph{w} & 12015 & -476 & -11837 & -113369\\
${\delta}u$ & -180070 & 214400 & 68340 & -56560\\
\emph{q} & -192085 & 180856 & 80177 & 182019\\
${\delta}s$ & 1097.63 & -684.42 & -220.27 & -742.08\\
  \hline
\end{tabular}
\caption{Work \emph{w} (J/kg) input and output, change in internal energy ${\delta}{u}$ \emph{W} (J/kg), heat \emph{u} (J/kg) input and output, and change in entropy ${\delta}s$ (J/kg$\cdot$K) for each stage of the cycle using Freon R-12 as the working fluid.}
\label{tb:tbWUQS_r12}
\end{center}
\end{table}

\begin{table}[h]
\begin{center}
\begin{tabular}{ | c || c | c | c | c |}
  \hline
Stage & 12 & 23 & 34 & 41\\
  \hline
\emph{w} & 16508 & -327 & -18904 & -145238\\
${\delta}u$ & -252416 & 258636 & 116530 & -26230\\
\emph{q} & -268924 & 190193 & 135434 & 199218\\
${\delta}s$ & 1563.51 & -764.79 & -403.08 & -856.66\\
  \hline
\end{tabular}
\caption{Work \emph{w} (J/kg) input and output, change in internal energy ${\delta}{u}$ \emph{W} (J/kg), heat \emph{u} (J/kg) input and output, and change in entropy ${\delta}s$ (J/kg$\cdot$K) for each stage of the cycle using Freon R-22 as the working fluid.}
\label{tb:tbWUQS_r22}
\end{center}
\end{table}

\begin{table}[h]
\begin{center}
\begin{tabular}{ | c || c | c | c | c |}
  \hline
Stage & 12 & 23 & 34 & 41\\
  \hline
\emph{w} & 13840 & -496 & -13674 & -153165\\
${\delta}u$ & -249558 & 302348 & 88920 & -69010\\
\emph{q} & -263398 & 251814 & 102594 & 243845\\
${\delta}s$ & 1549.4 & -979.67 & -289 & -1023.79\\
  \hline
\end{tabular}
\caption{Work \emph{w} (J/kg) input and output, change in internal energy ${\delta}{u}$ \emph{W} (J/kg), heat \emph{u} (J/kg) input and output, and change in entropy ${\delta}s$ (J/kg$\cdot$K) for each stage of the cycle using Freon R-134a as the working fluid.}
\label{tb:tbWUQS_r134a}
\end{center}
\end{table}

\clearpage


\section{The Supercritical Stirling Cycle Heat Engine}

A Stirling engine cycle is defined by isothermal compression at the cold sink (stage 1-2), isochoric heating from the cold to the hot temperature (stage 2-3), isothermal expansion at the hot source (stage 3-4), and isochoric cooling back from the hot temperature to the cold temperature (stage 4-1).  In order that the ideal gas Stirling engine achieve maximum efficiency, there must be perfect regeneration from the isochoric cooling to the isochoric heating.  This is thermodynamically possible (though difficult in practice) as the specific heat of an ideal gas is constant regardless of volume, and thus $Q_{23}=Q_{41}$ over the same temperature range.  Provided there is this perfect regeneration, $Q_{in}=Q_{34}$ and $Q_{out}=Q_{12}$.  For an ideal gas subject to the equation of state defined in equation \ref{eq:eqIdealGas}
\begin{eqnarray}
\label{eq:eqIdealGas}
{{P}{\cdot}{v}}&=&{{R}{\cdot}{T}},
\end{eqnarray}
undergoing isothermal expansion \cite{2}, the heat input $q_{{\delta}T=0}$ (J/kg) is equal to the work output $W_{{\delta}T=0}$ (J/kg) defined by equation \ref{eq:eqWdef}
\begin{eqnarray}
\label{eq:eqIdealGasIsothermal}
{q_{{\delta}T=0}}={W_{{\delta}T=0}}={\int}{P}{\cdot}{dv}={{R}{\cdot}{T}{\cdot}{\int}{\frac{dv}{v}}}=R{\cdot}T{\cdot}log(\frac{V_2}{V_1}),
\end{eqnarray}
and thus the efficiency of an ideal gas Stirling engine is
\begin{eqnarray}
\label{eq:eqIdealGasStirling}
\eta={1-(\frac{Q_{out}}{Q_{in}})}={1-(\frac{Q_{12}}{Q_{34}})}={1-(\frac{R{\cdot}{T_L}{\cdot}log(\frac{V_2}{V_1})}{R{\cdot}{T_H}{\cdot}log(\frac{V_2}{V_1})})}={1-(\frac{T_L}{T_H})},
\end{eqnarray}
which is the Carnot efficiency defined in equation \ref{eq:eqEfficC}.  

Equation \ref{eq:eqIdealGasIsothermal} no longer applies when a working fluid is no longer an ideal gas (equation \ref{eq:eqIdealGas}) but a real fluid subjected to intermolecular forces such as the Van der Waal forces.  Just like gravity has been observed to be an entropic force \cite{Entropy01, Entropy02, Entropy03, Entropy04, Entropy05, Entropy06}, it can be expected that this temperature-dependent attractive intermolecular force can impact the entropy generated as a result of thermodynamic processes involving real working fluids.  As a fluid gets more and more dense, the molecules get closer to each other, and the impact of intermolecular forces increases.  When a Stirling Engine uses a dense real fluid as its working fluid, the internal energy will in fact change during isothermal compression and expansion, which can be found with equation \ref{eq:eqdU_ideal}, and can be simply and accurately approximated with empirical equation \ref{eq:eqDeltaU_mytheory}.

The author proposes a Stirling engine, using supercritical argon gas as the working fluid.  The reduced specific volume at top and bottom dead center are $V_R=1.5$ and $V_R=15.0$, whereas the reduced specific temperatures are $T_R=1.2951$ and $T_R=1.9786$ at the low and hot temperature range.  Argon has a molar mass of 39.9 g/mole, a critical pressure of 4.863 MPa, a critical temperature of 150.687 K, and a critical specific volume of 1.8692 cm$^3$/g; therefore the temperature of this Stirling engine ranges between -78$^{\circ}$C and 25$^{\circ}$C, and the specific volume ranges between 2.80374 cm$^3$/g and 28.0374 cm$^3$/g.  The intermolecular attractive parameter \emph{a'} defined in equation \ref{eq:eqDeltaU_mytheory} is thus 1,063.8 {Pa}${\cdot}${K$^{0.5}$}$\cdot${{m$^6$}${\cdot}$kg$^{-2}$} for argon.  The pressures within this supercritical argon heat engine cycle can be obtained using the Peng-Robinson equation of state \cite{PR1976,PitzerAcentric}
\begin{eqnarray}
\label{eq:eqPengRobinson}
{P}&=&{\frac{R{\cdot}T}{V-B}}-{\frac{A{\cdot}{\alpha}}{{V^2}+{2{\cdot}B{\cdot}V}-{B^2}}},\\ \nonumber
A&=&0.45724{\cdot}{\frac{{R^2}{\cdot}{T_c^2}}{P_c}}, \\ \nonumber
B&=&0.07780{\cdot}{\frac{R{\cdot}{T_c}}{P_c}}, \\ \nonumber
{\alpha}&=&{({1+{{\kappa}{\cdot}{({1-{\sqrt{T_R}}})}}})^2}, \\ \nonumber
{\kappa}&=&{0.37464}+{1.54226{\cdot}{\omega}}-{0.26992{\cdot}{\omega^2}}, \nonumber
\end{eqnarray}
where $\omega$ is Pitzer's acentric factor, defined as
\begin{eqnarray}
\label{eq:eqPengRobinson}
{\omega}&=&log_{10}(\frac{P_c}{P_{S}'})-1,
\end{eqnarray}
where $P_{S}'$ (Pa) is the saturated pressure at a reduced temperature of $T_R=0.7$, and $P_c$ (Pa) is the critical pressure.  For all of the monatomic fluids including argon and xenon, ${\omega}=0$.

\begin{table}[h]
\begin{center}
\begin{tabular}{ | c || c | c | c || c | c | c |}
  \hline
Stage & \emph{P} (MPa) & \emph{v} (cm$^3$/g) & \emph{T} ($^{\circ}$C) & $P_R$ & $v_R$ & $T_R$\\
  \hline
1 & 1.3744 & 28.0374 & -78 & 0.28263 & 15 & 1.2951\\
  \hline
2 & 9.6737 & 2.80374 & -78 & 1.9893 & 1.5 & 1.2951\\
  \hline
3 & 20.6058 & 2.80374 & 25 & 4.2373 & 1.5 & 1.9786\\
  \hline
4 & 2.1744 & 28.0374 & 25 & 0.44713 & 15 & 1.9786\\
  \hline
\end{tabular}
\caption{Pressure \emph{P} (MPa), specific volume \emph{v} (cm$\mathrm{^3}$/g), temperature \emph{T} ($^\circ$C), reduced pressure $P_R$, reduced specific volume $v_R$, and reduced temperature $T_R$, for the Stirling cycle heat engine utilizing supercritical argon gas as the working fluid.}
\label{tb:tbPvEngine_Ar}
\end{center}
\end{table}

By integrating the pressure and the change in volume during the isothermal compression of stage 1-2 and the isothermal expansion of stage 3-4, the work
\begin{eqnarray}
W&=&P{\cdot}{dV}, \nonumber
\end{eqnarray}
input $W_{in}$ (J/kg) and output $W_{out}$ (J/kg) can be determined
\begin{eqnarray}
{W_{in}}&=&78,237, \nonumber \\ \nonumber
{W_{out}}&=&136,064,
\end{eqnarray}
and by using the proposed equation \ref{eq:eqDeltaU_mytheory} for the change in internal energy during isothermal compression ${{\delta}u_{12}}$ (J/kg) and isothermal expansion ${{\delta}u_{34}}$ (J/kg)
\begin{eqnarray}
{{\delta}u_{12}}&=&24,445, \nonumber \\ \nonumber
{{\delta}u_{34}}&=&19,777,
\end{eqnarray}
the isothermal heat output $Q_{12}$ (J/kg) and input $Q_{34}$ (J/kg) can be determined
\begin{eqnarray}
Q_{12}&={W_{in}}+{{\delta}u_{12}}=&102,682, \nonumber \\ \nonumber
Q_{34}&={W_{out}}+{{\delta}u_{34}}=&155,841. 
\end{eqnarray}
This engine assumes perfect regeneration, where all of the heat output from isochoric cooling $Q_{41}$ (J/kg) is used for isochoric heating $Q_{23}$ (J/kg).  This is extremely difficult to practically implement, but absolutely possible thermodynamically.  For an ideal gas $Q_{23}=Q_{41}$; for a real gas this is not the case.  In order to determine the difference in heat needed from the hot source ${\delta}Q_{23}$ (J/kg)
\begin{eqnarray}
{\delta}Q_{23}={Q_{23}}-{Q_{41}}={Q_{12}}-{Q_{34}}+{W_{out}}-{W_{in}}={{\delta}u_{12}}-{{\delta}u_{34}}=4,668, \nonumber
\end{eqnarray}
and this additional heating requirement can be used to find the heat input $Q_{in}$ (J/kg) and output $Q_{out}$ (J/kg) of this engine
\begin{eqnarray}
Q_{in}=Q_{34}+{{\delta}Q_{23}}=160,509, \nonumber \\ \nonumber
Q_{out}=Q_{12}=102,682.  
\end{eqnarray}
The heat input and output can be used to find the thermodynamic efficiency of this heat engine
\begin{eqnarray}
{\eta_{HE}}={1-{\frac{Q_{out}}{Q_{in}}}}={1-{\frac{102,682}{160,509}}}=36.027\%, \nonumber
\end{eqnarray}
which exceeds by 4.3\% the theoretical ideal-gas Stirling-cycle efficiency defined in equation \ref{eq:eqIdealGasStirling} 
\begin{eqnarray}
{\eta}={1-{\frac{T_{L}}{T_{H}}}}={1-{\frac{1.2951}{1.9786}}}=34.545\%. \nonumber
\end{eqnarray}

The change in specific entropy ${\delta}s$ (J/kg$\cdot$K) is defined in equation \ref{eq:eqSideal}, and the total entropy generated ${{\delta}s_u}$ (J/kg$\cdot$K) in the universe by this interally reversible cycle is thus
\begin{eqnarray}
{{\delta}s_u}&=&{\frac{Q_{out}}{T_L}}-{\frac{Q_{in}}{T_H}}, \nonumber \\
&=&{\frac{102,682}{195.15}}-{\frac{160,509}{298.15}},\nonumber \\
&=&-12.180, \nonumber
\end{eqnarray}
remarkably demonstrating a net reduction in entropy throughout the universe with a practical piston-cylinder heat engine that has a realistic compression ratio of 10.


\section{Conclusion}

The author demonstrated a theoretical heat engine cycle, utilizing experimentally validated thermodynamics tables for fourteen different saturated fluids that have been verified by NIST \cite{NIST_Webbook}, to show how Clausius Theorem (equation \ref{eq:eqClausius}) is an imperfect description for the second law of thermodynamics.  The NIST thermodynamic tables for saturated fluids used to demonstrate these claims have been utilized in research and industry for decades.  This was expanded to a practical Stirling engine utilizing supercritical argon as the working fluid.  In addition, equation \ref{eq:eqDeltaU_mytheory} represents a validated empirical equation for the change in internal energy of a real fluid undergoing isothermal compression or expansion.  This effort demonstrates how temperature dependent intermolecular attractive forces that are present in real fluids, including saturated gases, saturated liquids, and supercritical fluids, can serve as a negative entropic force, resulting in an decrease in net global entropy, and a macroscopic heat engine that exceeds the Carnot-limited thermodynamic efficiency!


\section*{Competing Interests}
\noindent The author declares no conflict of interest.

\section*{Acknowledgments}
\noindent The author would like to thank Mollie Marko and Michael Clark for useful discussions.


\bibliographystyle{unsrt}

\newpage


\end{document}